\documentclass[letterpaper,prl]{article}
\usepackage[dvips]{graphicx}
\begin{document}
\title{Stationary States of a Random Copying Mechanism over a Complex Networks}
\author{C\'esar A. Hidalgo R.$^{1}$$^{\dag}$, Francisco Claro H.$^1$ and Pablo A. Marquet$^2$\\
\scriptsize $^1$ Facultad de F\'isica. Pontificia Universidad
Cat\'olica de Chile.\\ \scriptsize Vicu\~na Mackenna 4860,
Mac\'ul, Santiago, Chile.\\ \scriptsize $^2$ Center for Advanced
Studies in Ecology and Biodiversity \& \\ \scriptsize Departamento
de
Ecolog\'ia. Pontificia Universidad Cat\'olica de Chile.\\
\scriptsize Casilla 114-D, Santiago, Chile\\ \scriptsize$^{\dag}$
To whom correspondence should be addressed.
\emph{chidalgo@nd.edu}}
\date{}
\maketitle
\begin{abstract}
An analytical approach to network dynamics is used to show that
when agents copy their state randomly the network arrives to a
stationary regime in which the distribution of states is
independent of the degree. The effects of network topology on the
process are characterized introducing a quantity called
\emph{influence} and studying its behavior for scale-free and
random networks. We show that for this model degree averaged
quantities are constant in time regardless of the number of states
involved.
\end{abstract}

\section{Introduction}
\indent Complex network theory has flourished as an effort to
explain the main characteristics of discrete interacting systems.
The first studies were able to describe and explain their topology
by showing that different thermodynamical quantities are needed to
characterize a particular network, such as its degree distribution
\cite{AlbertNature200,BarabasiScience1999,AmaralPNAS2000,WattsNature1998,AlbertReview2002},
community structures \cite{NewmanLetters2002,GirvanPNAS2002},
resilience \cite{NewmanReviewE2003} and motifs
\cite{MiloScience2002}. Recent efforts have also studied
quantities spreading on complex networks, such as computer viruses
\cite{Moreno,PastorSatorras2001,PSReviewE2002,Barabasi2002,NewmanReviewE}
in technological networks and information in
their social counterpart \cite{WuArxiv2003}.\\
\indent On the other hand, state dynamics has been studied
disguised as opinion formation processes, which have been modelled
using approaches including cellular automata
\cite{HolystarXiv2000}, spin chains
\cite{SznajdIJMPC2000,SlaninaEPJB2003} and opinion drifts over a
continuous opinion space \cite{LagunaarXiv2003}, without
incorporating topology. Recently, network topology has become more
relevant and computer simulations have been performed in this
direction
\cite{StaufferIJMPC2004,GonzalezIJMPC2004,BonnekohIJMPC2003},
though a systematic way to evaluate the topological effects on the
dynamical processes taking place on networks is still
missing.\\
\indent In this paper we study the effects of topology on a copy
mechanism taking place over a network with a non-trivial topology
and an arbitrary number of states. To achieve this, we group
agents with similar topological characteristics and study the
density of the evolving quantities over these groups by
considering
the interactions among them.\\
We will first introduce the model together with its associated
notation and then solve it analytically. We will then study how
network topology affects the simple dynamical process introduced
below and show that the system arrives to a configuration in which
the distribution of states is independent of the degree.
\section{The Model}
\indent We consider a process in which at each time step an agent
changes its state by copying one randomly from one of its
immediate neighbors. The probability that this occurs is the
product between the probability that the agent is chosen, times
the fraction of neighbors in a given state, including the agent,
that are acquainted by it. We will consider a directed network
with an adjustable topology as the substrate in which these
dynamical processes will occur, and attempt to characterize the
influence of topology upon them. We assume that changes in the
network topology take place at larger timescales than the
relaxation times of the system.\\
\subsection{Definitions}
\indent We begin by grouping agents according to their degree into
aggregates we call \emph{guilds}. This reduces the problem to one
as big as the number of guilds used to calculate the
approximation, this depends on the desired degree resolution, but
it is always considerable smaller than the number of agents.
Agents in a specific guild can be in any of the available
\emph{states}. We introduce the density $\rho_{i\alpha}$ as the
number of agents in the $i^{th}$ guild which pertain to the
$\alpha$ state, over the number of agents in the network. We
define \emph{guild size} $\rho_{i}=\sum_{\alpha}\rho_{i\alpha}$ as
the fraction of agents constituting that guild and
\emph{fractional abundance} as the fraction of agents on a
particular state in the entire network
$\rho_{\alpha}=\sum_{i}\rho_{i\alpha}$. From now on, latin indices
will refer to guilds, while greek indices will label
states.\\
\indent We assume that the linking patterns of agents in the same
guild are similar. In other words, we will model a network in
which agents from the same guild are linked to a similar number of
agents in other guilds, although not necessarily the same. This
motivates us to introduce $P_{ij}$ as the probability that an
agent in the $i^{th}$ guild is linked to an agent in the $j^{th}$
guild. We will refer to these quantities as \emph{linking
probabilities} and say that someone is linked to someone else if
the linking agent has a \emph{personal} or \emph{media} based
knowledge of the linked one.\\
\begin{figure}
\begin{center}
\includegraphics[width=0.5\textwidth]{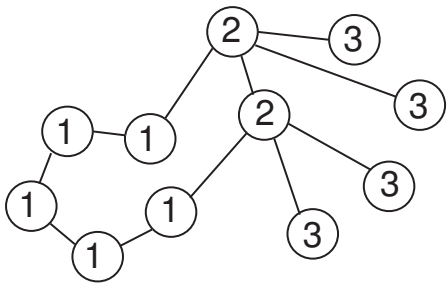}
\caption{\label{explain} Undirected network with eleven nodes in
which 3 guilds are distinguishable. Nodes carry guilds as its
labels, the first guild has 5 agents with a degree equal to 2. The
second guild is formed by 2 agents with a degree equal to 4. The
third guild is formed by 4 agents with a degree equal to 1.}
\end{center}
\end{figure}
An example is shown in figure (\ref{explain}). Three equal size
guilds are clearly distinguishable and acquaintanceship
probabilities are given by
$$
P=\left(
\begin{array}{ccc}
2/5 & 1/5 & 0 \\
1/5 & 1 & 1/2 \\
0 & 1/2 & 0 \\
\end{array}
\right).
$$
\subsection{Topologically Considerate Mean-Field Approach}
\indent In the large network limit the rate of change for a state
in a given guild is equal to the difference between the
probabilities of wining and losing an agent. The probability that
an agent in the $i^{th}$ guild turns into the $\alpha^{th}$ state
is
\begin{equation}
\label{P+}
\Pi_{i\alpha}^{+}=\frac{{\displaystyle\sum_{j}P_{ij}\rho_{j\alpha}\sum_{\mu\neq\alpha}\rho_{i\mu}}}{{\displaystyle\sum_{\mu,j}P_{ij}\rho_{j\mu}}},
\end{equation}
while the probability that an agent in the $i^{th}$ guild and the $\alpha^{th}$ state changes its state to a diffrenet one is
\begin{equation}
\label{P-}
\Pi_{i\alpha}^{-}=\frac{{\displaystyle\rho_{i\alpha}\sum_{\mu\neq\alpha,j}P_{ij}\rho_{j\mu}}}{{\displaystyle\sum_{\mu,j}P_{ij}\rho_{j\mu}}}.
\end{equation}
\indent We simplify the notation introducing $\Upsilon$ as the
matrix representing the expected fraction of
agents in a particular state known by an agent in a particular guild on the entire network,\\
\begin{equation}
\label{Upsilon} \Upsilon_{i\alpha}=\sum_{j}P_{ij}\rho_{j\alpha}.
\end{equation}
\indent Combining eqns. (\ref{P+}), (\ref{P-}) and
(\ref{Upsilon}) we can write the temporal variation of state
abundance for a given guild as
\begin{equation}
\label{Master}
\partial_{t}\rho_{i\alpha}=\frac{1}{\Upsilon_{i}}\bigg(
\Upsilon_{i\alpha}\sum_{\mu}\rho_{i\mu}-\rho_{i\alpha}\sum_{\mu}\Upsilon_{i\mu}\bigg),
\end{equation}
where $\Upsilon_{i}=\sum_{\mu}\Upsilon_{i\mu}$ represents the
fraction of agents acquainted by the $i^{th}$ guild\footnote{From
now on we will assume that quantities with a missing index have
been added over it, for example:
$\rho_{\alpha}=\sum_{i}\rho_{i\alpha}$ }. The extension of the
restricted sums in equations (\ref{P+}) and (\ref{P-}) to include
all states has no effect on the system dynamics because the
additional terms cancel out when we place them on eqn.
(\ref{Master}). Time units represent the network's natural update
time, which is the number of time steps equal to the population
size in a random or sequential
updating processes.\\
\subsection{Analytical Solution}
\indent An analytical solution can be obtained by re-writing eqn.
(\ref{Master}) and performing the substitution
$\rho_{i\alpha}=U_{i\alpha}e^{-t}$. This transforms eqn.
(\ref{Master}) into an eigenvalue equation for the vector
$\vec{U_{\alpha}}=(U_{1\alpha}, U_{2\alpha}, \cdots)$. The problem
then reduces to finding the solution of
\begin{equation}
\label{eigenMaster}
\partial_{t}\vec{U_{\alpha}}=M\vec{U_{\alpha}},
\label{M}
\end{equation}
where $M_{ij}=\frac{\rho_{i}}{\Upsilon_{i}}P_{ij}$. From eqn.
(\ref{eigenMaster}) it follows that the eigenfunctions of $M$
depend exponentially on time with a decay rate $\lambda$, where
$\lambda$ is the associated eigenvalue, this agrees with the
results presented in \cite{Wu2004,Sucheki2004}. A general solution
can be written in terms of this basis as
\begin{equation}
\label{Solution}
\rho_{i\alpha}=\sum_{\eta}C^{\eta}_{i\alpha}V^{\eta}_{i}e^{(\lambda_{\eta}-1)t},
\end{equation}
in which $V^{\eta}_{i}$ is the $i^{th}$ element of the $\eta^{th}$
eigenvector of the matrix $M$ and $\lambda_{\eta}$ is its associated
eigenvalue. It is important to notice that $M$ does not depend on the state, implying
that the state dependence is introduced through the initial
conditions only, represented by $C_{i\alpha}^{\eta}$.\\
\subsubsection{System Equilibria}
\begin{figure}[ht]
\begin{center}
\includegraphics[width=0.6\textwidth]{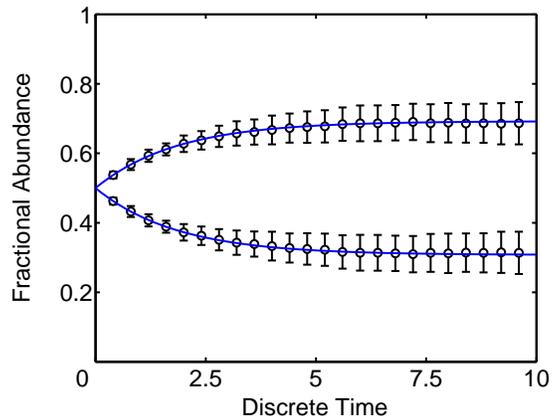}
\caption{\label{stoch}The continuous line represents the mean-field
solution of a two guild network with 1000 agents, while circles
represent an average of 50 stochastic simulations. Error bars
indicate one standard deviation. In each simulation the network was
chosen randomly constrained to satisfy the conditions presented
in the text.}
\end{center}
\end{figure}

\indent The steady states are characterized by the eigenvector
associated with the eigenvalue $\lambda=1$. By direct calculation
one can show that $M$ always has an eigenvalue equal to 1
associated with guild sizes as its eigenvector. One has
$$
\sum_{j}M_{ij}\rho_{j}=\sum_{j}\frac{\rho_{i}P_{ij}
\rho_{j}}{\Upsilon_{i}}=\rho_{i}\sum_{j}\frac{P_{ij}\rho_{j}}{\Upsilon_{i}}=\rho_{i}.
$$
We have observed numerically that the remaining
eigenvalues have a real part smaller than 1 and lie inside the
unit circle of the complex plane, a formal demonstration of this remains open.\\
\indent The mean-field approach presented above was compared with
a stochastic simulation on a 1000 agents network in which two
equal size guilds were distinguishable (fig. (\ref{stoch})). The
acquaintanceship probabilities were chosen such that all agents
acquainted everyone in their respective guild, while the ones in
the first guild acquainted an average of 3/5 of the agents in the
second guild. On the contrary, agents in the second guild
acquainted 1/5 of the agents in the first one. The simulation
shows how accurate the mean field model is and reveals that the
standard deviation tends to stabilize as
the system relaxes to equilibrium.\\
\section{Degree Distribution}
\indent We define the $in$ degree of an agent pertaining to a
specific guild as the fraction of agents in the total sample that
it knows. It, therefore represents the fraction of the network
exerting a direct influence on it. On the other hand, an agent's
$out$ degree is the fraction of the entire network that knows the
agent, these are the ones directly influenced by it. This
definition of degree implies that peer influence flows opposite to
acquaintanceship. According to this, the $in$ and $out$ degree's
in our model are represented by
\begin{equation}
\label{k} k_{i}^{in}=N\sum_{j}P_{ij}\rho_{j}=N\Upsilon_{i}\:,\:\quad
k_{i}^{out}=N\sum_{j}P_{ji}\rho_{j}\:,
\end{equation}
thus degree distribution can be imposed by considering that the fraction
of agents with the described degree is equal to guild size
$\rho_{i}$. We can approach a complex network with an \emph{out} degree
distribution
satisfying a function $f$ by the solutions of
\begin{equation}
\label{ForceNet} \rho_{i}=f\bigg(N\sum_{j}P_{ji}\rho_{j}\bigg)
\end{equation}
\indent In the case of scale-free networks eqn. (\ref{ForceNet})
becomes
\begin{equation}
\label{sfeq} \rho_{i}=\bigg(\sum_{j}P_{ji}\rho_{j}\bigg)^{-\gamma},
\end{equation}
where constant factors like normalization or network size have been
absorbed by the elements of $P$, noting that equation
(\ref{Master}) is zero degree homogeneous in $P$. For the case of
a random network, degree distribution has an
exponential tail that can be approximated by the solutions of
\begin{equation}
\label{sweq} \rho_{i}=A\exp{\bigg(-\eta\sum_{j}P_{ji}\rho_{j}\bigg)}
\end{equation}
as parameters. Here $\eta$ characterizes the rate of exponential
decay and $A$ is a normalization factor.
\section{Guild Influence}
\subsection{Scale-Free Networks} \indent To capture
the influence exerted by the topology on the dynamical process, we
define \emph{guild influence} $I_{i}$ as the equilibrium
fractional abundance that a state will achieve assuming that it
was initially occupying, only and entirely, the $i^{th}$ guild
($\rho_{\alpha}(t=0)=\rho_{i}$),
$$
I_{i}=\lim_{t\rightarrow\infty}\frac{\rho_{\alpha}(t)}{\rho_{i}}.
$$
\indent This quantity represents the average number of copies that
an agent of a given guild creates during the entire process,  or
an effective infection rate proper of the guild. This is a
formalization of what Wu et. al.\cite{Wu2004} and Sucheki et.
al.\cite{Wu2004} have done. Influence measures the affect of the
group of agents with a certain degree,
generalizing the bias depicted in the two state systems studied by them.\\
\begin{figure}[htb]
\begin{center}
\includegraphics[width=0.7\textwidth]{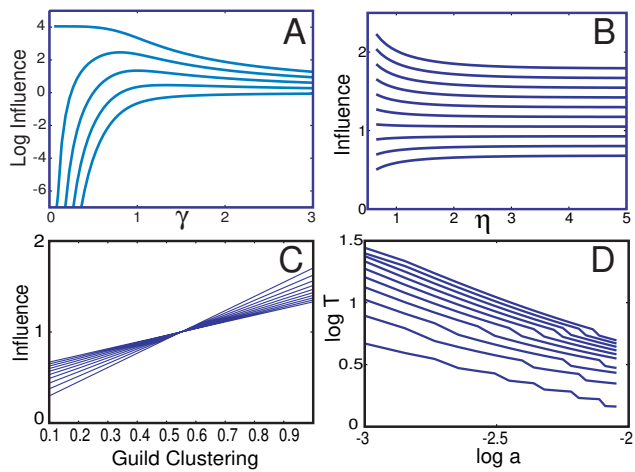}
\caption{\label{sf} (A) Guild influence against scale-free exponent
$\gamma$ for a 5 guild network calculated through numerical
integration of Eq. (\ref{Master}). From top to bottom, guild sizes
were chosen as $9\times10^{-1}$, $9\times10^{-2}$, $9\times10^{-3}$,
$9\times10^{-4}$, and $9\times10^{-5}$. (B)Influence structure of a
random network against the inverse of its characteristic length
$\eta$, for a 10 guild network. From top to bottom guild sizes were
chosen as 0.026, 0.032, 0.041, 0.053, 0.068, 0.088, 0.114, 0.147,
0.189 and 0.245. (C) Influence for a ten guild network in which each
guild had a particular clustering coefficient. Different lines
represent linearly spaced variations of the constant $a$ in the
0.01-0.09 interval from lowest to highest slope respectively. The
constant $b$ was taken as 0.1 and 10 guilds were used. (D)
Characteristic times against inter-guild connectivity for values of
$b$ chosen linearly spaced from top to bottom in the 0.1-0.01
interval }
\end{center}
\end{figure}
\indent We studied guild influence on a scale-free network by
choosing guild sizes and finding acquaintanceships probabilities
for a given scale-free exponent according to eqn. (\ref{sfeq}).
Figure (\ref{sf}A) shows that the influence of the highest degree
guild increases with decreasing $\gamma$, until it reaches a
steady value which is equal to the ratio between the system size
and the number of agents in the highest degree guild. This is a
consequence of the increasing fame of the highest degree guild
that occurs for small values of $\gamma$. In the large exponent
limit the influence of all guilds approach unity, meaning that
there are no high
\emph{out} degree agents exerting an important influence on the system.\\
\subsubsection{Exponential Networks}
\indent An example involving a exponential network approximated by
10 guilds is shown in fig. (\ref{sf}B), the procedure used to
determine the parameters being analogous to the one performed for
the scale-free network. The figure shows that in a random network
high degree guilds exert a mild influence compared to that
occurring on its scale-free counterpart. This is a consequence of
the reduced
population of high degree guilds.\\
\subsection{Clustering} The topological influence of clustering was
studied by choosing acquaintanceship probabilities as
$$
P_{ij}=a+\delta_{ij}(b\cdot i-a) \quad \textrm{with} \quad b>a
$$
and equal guild sizes for all the network, where $\delta_{ij}$ is
the Kronecker delta. This model represents a random network in
which agents of the $i^{th}$ guild have a probability $a$ of being
acquainted to agents in other guilds. As usual, we shall call this
connectivity. On the contrary, the probability $b\cdot i$ to
acquaint someone from its own guild together with the condition
$b>a$ defines clustered groups with an internal connectivity that
increases linearly throughout the network. This is similar to
communities in the sense introduced by Newman and
Girvan\cite{GirvanPNAS2002,Glesier2003} and are also similar to
the networks studied by Watts\cite{SmallWorlds} at the end of last
century . A particular
case of this networks is the one shown in figure (\ref{explain}).\\
\indent In a random network, clustering is equal to the
probability that two agents are acquainted, motivating us to
define guild clustering as the clustering coefficient of a guild
in the absence of the rest of the network. We studied the
dependence of influence on guild clustering (fig. (\ref{sf}C))
finding that there is a linear dependence with a slope that
decreases with $a$, meaning that the reinforcement mechanism is
more effective in sparsely connected networks than in strongly
connected ones. Either way, the effect continues to remain mild
compared to its scale-free counterpart. On the other hand, the
characteristic time $T$ required for the system to converge
increases with decreasing $a$ as $T\sim
a^{-\kappa(b)}$ with $d\kappa/db\: <\: 0$ (see Fig. \ref{sf}D).\\

\section{Equilibrium Distribution}

Using the formalism recently introduced we can also find how
states are distributed when the network reaches equilibria. In
order to do this, first we notice that eqn.(\ref{Master}) vanishes when
\begin{equation}
\label{condition1}
\frac{\Upsilon_{i\alpha}}{\Upsilon_{i}}=\frac{\rho_{i\alpha}}{\rho_{i}}.
\end{equation}
If we multiply eqn. (\ref{Master}) by $P_{ji}$ and add over all
$i$ we will find a rate eqn. for $\Upsilon_{j\alpha}$. This, must
vanish in equilibria implying that
\begin{equation}
\label{condition2}
\Upsilon_{j}\sum_{i}\frac{\Upsilon_{i\alpha}}{\Upsilon_{i}}=\sum_{i}\Upsilon_{j\alpha}.
\end{equation}
The term on the righthand side does not depend on $i$ so the
summation just yields the number of guilds we have chosen to
approximate the system. Considering a system with $m$ guilds leads
us to conclude that (\ref{condition2}) is the same as
\begin{equation}
\label{condition2f}
\frac{1}{m}\sum_{i}\frac{\Upsilon_{i\alpha}}{\Upsilon_{i}}=\frac{\Upsilon_{j\alpha}}{\Upsilon_{j}}.
\end{equation}
The lefthand side of this equality does not depend on $j$, so when
the system reaches the steady state, the ratio between
$\Upsilon_{j\alpha}/\Upsilon_{j}$ must be the same for every
guild. Combining this with condition (\ref{condition1}) we
conclude that upon equilibria the system reaches a configuration
in which the ratio between the number of agents in a certain state
over the total number of agents is independent of the degree, thus
each guild has exactly the same distribution of states as the
entire network.\\

\section{Network Structure}

The formalism presented above can be used to understand some
topological characteristics of the studied network. Recently, two
groups have presented works indicating that a simple dynamical
process like the one studied in this paper can be solved, in the
case in which two states are involved, by the use of conserved
quantities or martingales \cite{Wu2004,Sucheki2004}. The first of
them proposes that a weighted average over degree should be
conserved in a dynamical process of this kind. In our formalism we
average the fraction of agents in each state weighted by their
relative degree as
\begin{equation}
\sum_{i\alpha}\Upsilon_{i\alpha}.
\end{equation}
We can shows that this is a conserved quantity by taking eqn.
(\ref{Master}), multiply it by $P_{ji}$ and add it over $i$, $j$
and $\alpha$ to get,
$$
\partial_{t}\Upsilon=\sum_{ij\mu\alpha}\bigg(\frac{P_{ji}\Upsilon_{i\alpha}\rho_{i\mu}}{\Upsilon_{i}}-\frac{\rho_{i\alpha}\Upsilon_{i\mu}P_{ji}}{\Upsilon_{i}}\bigg).
$$
Performing the summation over $\alpha$ on the first term and over
$\mu$ in the second term and cancelling the $\Upsilon_{i}$ we get.
$$
\partial_{t}\Upsilon=\sum_{ij\mu}P_{ji}\rho_{i\mu}-\sum{ij\alpha}P_{ji}\rho_{i\alpha}
$$
which is clearly equal to zero. This quantity is naturally
conserved if we consider that $\Upsilon_{i\alpha}$ represents the
fraction of agents in the $\alpha$ state that are known by someone
on the the $i'th$ guild. Thus $\Upsilon_{i}$ represents the total
numbers of agents know by the $i'th$ guild, and $\Upsilon$
therefore represents the fraction of agents known by all the
guilds combined, or in the other words the number of
non-intersecting sub-graphs of the network that are able to reach
all the nodes of it.\\
In a two state system in which each one of them is represented by
a spin up or down we can show that the degree averaged
magnetization does not depend on time by replacing the fraction of
agents in a certain state by the fraction of agents in that state
times its spin.
$$
\rho_{i\alpha}\rightarrow\rho_{i\alpha}\sigma_{\alpha}
$$
If we span $\Upsilon_{i\alpha}$ we can show that the change on
this quantity can be constructed by using our formalism as
$$
\partial_{t}\sum_{ij\alpha}P_{ij}\rho_{j\alpha}\sigma_\alpha=\sum_{ij\mu\alpha k}\bigg(\frac{P_{ji}P_{ik}\rho_{k\alpha}\sigma_{\alpha}\rho_{i\mu}\sigma_{\mu}}{\Upsilon_{i}}-\frac{\rho_{i\alpha}P_{ik}\rho_{k\mu}P_{ji}}{\Upsilon_{i}}\bigg).
$$
where the two added terms are strictly equal if we consider that
$\mu$ and $\alpha$ are mute variables. This agrees with the
results found by Wu et. al. \cite{Wu2004} and can be extended
to an arbitrary number of spins. Which in this case are nothing more
than labels, and we believe its \emph{names}
should not enter the calculations.\\

\section{Conclusion}

Using a topologically considerate mean-field approach we were able
to characterize the simple dynamics of a node updating mechanism
on a complex network with an arbitrary degree distribution.\\
The main idea behind this technique was to group nodes with
similar topological characteristics into aggregates we called
guilds. We characterized the reproductive capability of guilds by
a quantity we called influence and showed that in the case of a
scale-free network the system bias dramatically towards the states
pertaining to the highest degree guilds as the characteristic
exponent decrease, while on an exponential network this effect is
comparatively mild.\\
It was also shown that the convergence time scale as a function of
connectivity, with a characteristic exponent close to 1, when we
considered a network formed by mildly connected clusters. This is a
consequence of the reinforcement mechanism introduced by the node
updating rule. We were also able to show that clustered groups
tend to be more influential than un-clustered ones. This is due to
the fact that they are harder to invade, allowing them to act over
the
system for a longer period of time.\\
The final configuration of the system was unravelled, showing that
it arrives to a stationary distribution in which the fraction of
nodes with a certain degree that are in a certain state, over the
total number of nodes with that degree, is independent of the
particular degree and it therefore mimics the network.\\
In the last section we discussed a purely topological quantity,
which is naturally conserved and  represents the number of
subgraphs present in the network with the ability to held it
together. This suggests a direct link with network resilience and
will be present in future discussions. We also showed that the
degree weighted average of states over the system remains
constant, regardless of the value and number of them in agreement
with previous results.\\

Work supported in part by Fondecyt project 1020829. PAM
acknowledges the support of the Santa Fe Institute through and
International Fellowship. CAH acknowledges the support of the
Helen Kellog Institute for International Studies at the University
of Notre Dame and to Professor Barab\'asi for discussing the paper
with us.

\end{document}